\documentclass[10pt,emptycopyrightspace]{ewsn-proc}

\usepackage{graphicx}
\usepackage{balance}
\usepackage{comment}
\usepackage{color}
\usepackage{graphicx}
\usepackage[font=small]{subfig}
\usepackage{url}
\usepackage{hyperref}
\usepackage[english]{babel}
\usepackage[utf8]{inputenc}
\usepackage{amsmath,amssymb}
\usepackage{algorithm}
\usepackage{algpseudocode}
\usepackage{multirow}

\usepackage{blindtext}
\newcommand{\etal}{et~al.\xspace}
\newcommand{\eg}{e.g.\xspace}
\newcommand{\ie}{i.e.\xspace}
\newcommand{\lr}{\mbox{LR-FHSS}\xspace}

\numberofauthors{3}
\author{
\alignauthor Juan A. Fraire\\
    \affaddr{Univ Lyon, Inria, INSA Lyon, CITI, EA3720, 69621 Villeurbanne, France\\
    Saarland University, Saarland Informatics Campus E1\,3, 66123 Saarbrücken, Germany}
    \email{juan.fraire@inria.fr}
\alignauthor Alexandre Guitton\\
    \affaddr{Univ Lyon, Inria, INSA Lyon, CITI, EA3720, 69621 Villeurbanne, France\\
    Université Clermont-Auvergne, CNRS, Mines de Saint-Étienne, Clermont-Auvergne-INP, LIMOS, 63000 Clermont-Ferrand, France}
    \email{alexandre.guitton@uca.fr}
\alignauthor Oana Iova\\
    \affaddr{Univ Lyon, INSA Lyon, Inria, CITI, EA3720, 69621 Villeurbanne, France\\}
    \email{oana.iova@inria.fr}
}

\title{Recovering Headerless Frames in LR-FHSS}

\begin{document}

\maketitle

\begin{abstract}
Long-Range Frequency Hopping Spread Spectrum (LR-FHSS) is a recent modulation designed for communications from low-power ground end-devices to Low-Earth Orbit (LEO) satellites. To decode a frame, an LR-FHSS gateway must receive at least one header replica and a large proportion of payload fragments. However, LR-FHSS headers will likely be lost when there are many concurrent transmissions. In this paper, we motivate the header loss problem with an analytical model, propose a linear programming model to extract headerless frames and design a cost-effective sliding-window heuristic. Simulation results show that our approach exhibits near-optimal headerless detection and extraction results while ensuring a low computational cost. The proposed method is, therefore, suitable for future LR-FHSS gateways located onboard resource-constrained LEO satellites.
\end{abstract}

%
% NOTE
%
% Do not provide category, terms, keywords for the reviewed submission.
% They will only be added for the camera-ready version. Instructions will
% be provided for the camera ready version.

%A category including the fourth, optional field follows...
% \category{D.2.8}{Software Engineering}{Metrics}[complexity measures, performance measures]
% \terms{Delphi theory}
% \keywords{ACM proceedings, \LaTeX, text tagging}

\section{Introduction}

Low-Earth Orbit (LEO) satellites enable a large variety of promising monitoring applications (e.g., environmental or ecological monitoring of oceans, poles, vast forests, extensive natural parks~\cite{desanctis16satellite}) with a cost significantly reduced compared to geostationary satellites. Such applications are otherwise difficult or impossible to serve with terrestrial infrastructure due to their geographical location~\cite{fraire2022space}.

Semtech recently introduced a new modulation called Long-Range Frequency Hopping Spread Spectrum (\lr)~\cite{lorawan-rp2103}, specifically designed for communications from low-power end-devices on the ground to LEO satellites. This modulation is supported by existing LoRaWAN gateways with a simple firmware update and works alongside legacy LoRa modulation, which will ease its adoption. Contrary to legacy LoRa, \lr is based on frequency hopping within several narrow-band sub-channels (of 488~Hz bandwidth), where the payload is divided into multiple short fragments, each sent on random sub-channels. The index of the chosen frequency hopping sequence is encoded in the header of the \lr frame, which is sent multiple times on random sub-channels in order to improve its robustness.  An \lr frame is received if at least one header replica is received, as well as a large proportion of the fragments (typically one-third or two-thirds). Thanks to the combination of narrow-band sub-channels and robust coding rates, a LoRaWAN gateway hosted onboard an LEO satellite can decode hundreds of simultaneous \lr transmissions, which fits the requirements of direct-to-satellite IoT communications~\cite{fraire2019direct}. 

However, recent research showed that header loss is a significant cause of \lr frame loss~\cite{ullah22analysis,maleki2022d2d}. This is due to both the small number of header replicas (typically 2 or 3) and the relatively larger time-on-air of the header (233~ms) compared to the time-on-air of a fragment (102.4~ms). Thus, \textbf{addressing the recovery of headerless \lr frames is crucial to improve the reliability of \lr communication.}

We propose in this paper a method to recover headerless \lr frames.
To our knowledge, this is the first method of its kind.
The core idea is to deduce the random sequence used by the fragments of a given frame.
To do this, we assume that it is possible to detect whether a sub-channel is busy or not at a given time slot. This knowledge enables us to detect the sequence that matches the observed channel occupancy among all frequency hopping sequences since all possible sequences are known by the LoRaWAN gateway.

Our specific contributions are:
\begin{enumerate}
    \item A mathematical derivation of the cause of \lr frame loss that motivates the recovery of header loss frames; 
    \item A formalization of the sequence detection problem as an integer linear program (ILP) model;
    \item A low-cost heuristic to recover headerless \lr frames leveraging resource-constrained processors.
\end{enumerate}
Finally, by means of an extensive simulation campaign, we present compelling evidence of the efficiency and effectiveness of the proposed headerless decoding approach for \lr. 

\begin{figure*}
    \centering
    \includegraphics[width=\textwidth]{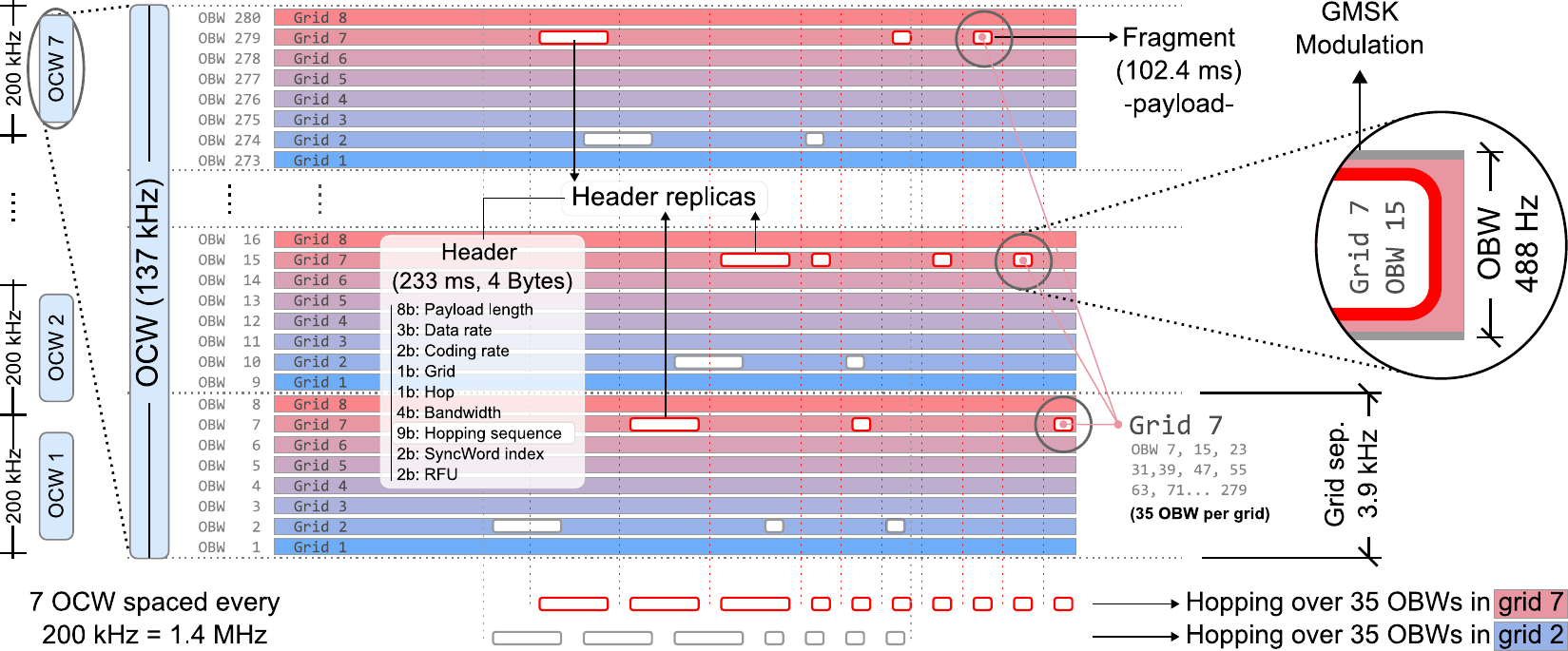}
    \caption{In \lr, the European 868~MHz frequency is divided into 7 channels called Occupied Channel Width (OCWs). Each OCW is divided into 280 sub-channels called Occupied Bandwidth (OBWs). To ensure that the mandatory minimum separation of 3.9kHz between two consecutive sub-channels, the notion of a grid has been introduced. Each OCW is divided into 8 grids, each grouping 35 OBWs (where two consecutive OBWs are separated by 3.9kHz). A frequency hopping sequence can use only OBWs inside a given grid.}
    \label{fig:channels}
\end{figure*}

The remainder of this paper is organized as follows. Section~\ref{section:state-of-the-art} introduces the \lr modulation as well as recent related works. Section~\ref{section:motivation} motivates our work through a probabilistic analysis that exposes the vulnerability of \lr headers. Section~\ref{section:proposal} first presents an optimal ILP to recover headerless frames, followed by a low-cost heuristic. Section~\ref{section:simulation-results} compares the results of the ILP and the heuristic through extensive simulations. Section~\ref{section:discussion} raises a discussion on our assumptions and on the mitigation of the false positives produced by our algorithms. Finally, Section~\ref{section:conclusion} concludes our work.

\section{\lr in a Nutshell}
\label{section:state-of-the-art}
\lr is a frequency hopping modulation that enables fast intra-frame frequency hopping with similar long-range capabilities as LoRa. It is one of the underlying modulations supported by LoRaWAN-compliant devices, together with LoRa and FSK~\cite{lorawan-rp2103}. While \lr is used for uplink communications (when energy-constrained end-devices  communicate with gateways located in LEO satellites), LoRa continues to be used for downlink communication. Indeed, with a simple firmware update, LoRaWAN gateways can decode  \lr frames without disturbing the reception and decoding of LoRa frames. In the reference gateway design, a gateway can listen to all available sub-channels at the same time. Its system capacity for decoding \lr frames is limited only by its digital signal processing (DSP) capacity~\cite{lrfhss-system}. The real-world limit is 700k packets per day with the current software. 

\lr uses Gaussian Minimum-Shift Keying (GMSK) modulation~\cite{lebreton22crash} in each sub-channel, which is highly spectral-efficient and energy-efficient. GMSK encodes the data by discrete frequency changes. The use of a Gaussian filter reduces the sidebands caused by the modulation, which is desired for narrow-band communications. GMSK is famous for being used in Global System for Mobile (GSM) communications, as well as for satellite communications.

To increase the recovery of frames in dense environments that are prone to collisions and interference, \lr also has built-in error correction mechanisms. The used coding rate (CR) is 1/3 or 2/3, depending on the data rate. 

\begin{table*}[]
\centering
\caption{Regional parameters for \lr OCWs, OBWs, and grids.}
\label{tab:reg_params}
\begin{tabular}{|c|c|c|c|c|c|c|c|c|c|}
\hline
Region &
\begin{tabular}[c]{@{}c@{}}OCW band-\\ width (MHz)\end{tabular} &
\begin{tabular}[c]{@{}c@{}}OBW band-\\ width (Hz)\end{tabular} &
\begin{tabular}[c]{@{}c@{}}Grid sep.\\ (kHz)\end{tabular} &
\begin{tabular}[c]{@{}c@{}}No. of \\ grids\end{tabular} &
\begin{tabular}[c]{@{}c@{}}No. of OBWs \\ per grid\end{tabular} &
\begin{tabular}[c]{@{}c@{}}No. of \\ OBWs\end{tabular} &
\begin{tabular}[c]{@{}c@{}}Coding\\ rate (CR)\end{tabular} &
\begin{tabular}[c]{@{}c@{}}Bit rate\\ (bits/s)\end{tabular} &
\begin{tabular}[c]{@{}c@{}}Data\\ rate\end{tabular} \\
\hline
\multirow{4}{*}{EU} & \multirow{2}{*}{137} & \multirow{2}{*}{488} & \multirow{2}{*}{3.9} & \multirow{2}{*}{8} & \multirow{2}{*}{35} & \multirow{2}{*}{\begin{tabular}[c]{@{}c@{}}280\\ (=8x35)\end{tabular}} & 1/3 & 162 & DR8 \\
\cline{8-10}
& & & & & & & 2/3 & 325 & DR9 \\
\cline{2-10}
& \multirow{2}{*}{366} & \multirow{2}{*}{488} & \multirow{2}{*}{3.9} & \multirow{2}{*}{8} & \multirow{2}{*}{86} & \multirow{2}{*}{\begin{tabular}[c]{@{}c@{}}688\\ (=8x86)\end{tabular}} & 1/3 & 162 & DR10 \\
\cline{8-10}
& & & & & & & 2/3 & 325 & DR11 \\
\hline
\multirow{2}{*}{US} & \multirow{2}{*}{1523} & \multirow{2}{*}{488} & \multirow{2}{*}{25.4} & \multirow{2}{*}{52} & \multirow{2}{*}{60} & \multirow{2}{*}{\begin{tabular}[c]{@{}c@{}}3125\\(=52x60)\end{tabular}} & 1/3 & 162 & DR5 \\
\cline{8-10} 
& & & & & & & 2/3 & 325 & DR6 \\
\hline
\end{tabular}
\end{table*}

In \lr, the frequency band is split into several contiguous Occupied Channel Width (OCWs).
% , inside which the frequency hopping for a given frame happens. 
These OCWs are then split into multiple Occupied Bandwidth (OBWs)~\cite{boquet21lrfhss}, where each OBW corresponds to a physical sub-carrier with a 488~Hz bandwidth. To enforce the minimum distance between sub-carriers used in the frequency hopping sequence, as imposed by regional regulations, the OBWs are organized into grids. 
% For example, in the 868 MHz frequency band used in Europe, the minimum frequency shift is 3.9 kHz for each fragment (see Table~\ref{tab:reg_params})~\cite{TBA}. 
Grids are a fundamental concept in \lr, as a hopping sequence of a given frame can only use OBWs corresponding to the same grid. Grids are groups of non-consecutive OBWs separated by a minimum distance, depending on the regional regulation. The number of OCWs, OBWs, and grids for the EU and the US is shown in Table~\ref{tab:reg_params}.

We exemplify the use of \lr in Europe in Figure~\ref{fig:channels}, following regional regulations. The European 868~MHz band is divided into 7 OCWs of 137~kHz, each containing 280 OBWs of 488~Hz. In Europe, the minimum separation between OBWs during the frequency hopping is 3.9~kHz; thus, the set of 280 OBWs is organized into 8 grids of 35 OBWs each. This ensures that every two OBWs inside the grid are separated by a multiple of 3.9 kHz. 
For example, grid 1 contains OBWs 1, 9, 17, ... , 273 (\ie, all the blue-colored OBWs). 

The transmission of a frame starts by sending several replicas of the header, followed by the payload divided into short fragments, each sent on a different OBW inside a grid. For example, we can see in Figure~\ref{fig:channels} that all the headers and the fragments of the red frame are transmitted only in pink-colored OBWs, which are part of grid 7. 
There are a total of $2^9=512$ possible hopping frequencies, known in advance by both end devices and gateways. Each replica of the header contains the index of the chosen frequency hopping sequence. To ensure robustness at reception, each replica is repeated two or three times (depending on the data rate scheme). The large number of possible sequences, which is $512$, makes it unlikely to have two or more frames share the same hopping sequence. The payload is encoded such that it can still be recovered even if a large proportion of the fragments are lost~\cite{boquet21lrfhss}. The two configurations specified in the LoRaWAN standard~\cite{lorawan-rp2103} are:
\begin{itemize}
    \item{}Robust configuration: each header is sent 3 times, and the payload is encoded using CR = 1/3 (at least 33\% fragments must be decoded to recover such a frame).
    \item{}Fast configuration: each header is sent 2 times, and the payload is encoded using CR = 2/3 (at least 67\% fragments must be decoded to recover such a frame).
\end{itemize}

With \lr, the gateway located onboard the LEO satellite constantly monitors all OBWs (for all OCWs) in order to detect header replicas. Once a replica is found, the gateway locks onto this transmission, extracts the index of the hopping sequence and the number of fragments from the replica, and follows the sequence in order to obtain the fragments. Since the gateway is mains powered, it is constantly listening for incoming frames. 

The novelty of the \lr technology results in a relatively sparse state of the art, with results obtained  analytically or in simulation.
Boquet~\etal~\cite{boquet21lrfhss} were the first to evaluate the performance of \lr, showing its considerable improvement in network capacity (up to two orders of magnitude) when compared to LoRa.
Ullah~\etal~\cite{ullah22analysis} showed through simulation that the primary reason for frame loss in \lr is the loss of the headers. Maleki~\etal~\cite{maleki2022d2d} evaluated the outage probability in LR-FHSS analytically with a realistic channel model.
This work was later extended in~\cite{maleki2022outage} to study the outage probability in direct-to-satellite-IoT.

As we can see, there is a lack of experimental work that shows how \lr works in practice, and how its communication is impacted by the environment. While we have a clear view of the effect of interference and collisions in legacy LoRa (\eg, non-orthogonality of spreading factors~\cite{croce18impact, benkhelifa22sf}, capture effect~\cite{noreen18lora}), we lack this information for \lr communication.

\section{Motivation: Fragility of \lr Headers}
\label{section:motivation}

In the following, we analytically compare the cause of frame loss in \lr, which can come from both header or payload loss. We consider the robust configuration (3 header replicas, CR=1/3), although the translation to the fast configuration is straightforward.

\begin{figure*}[htbp]
    \centering
    % % trim={<left> <lower> <right> <upper>}
    \subfloat[]{\includegraphics[width=.49\textwidth,trim={0.6cm 0.6cm 1cm 0cm},clip]{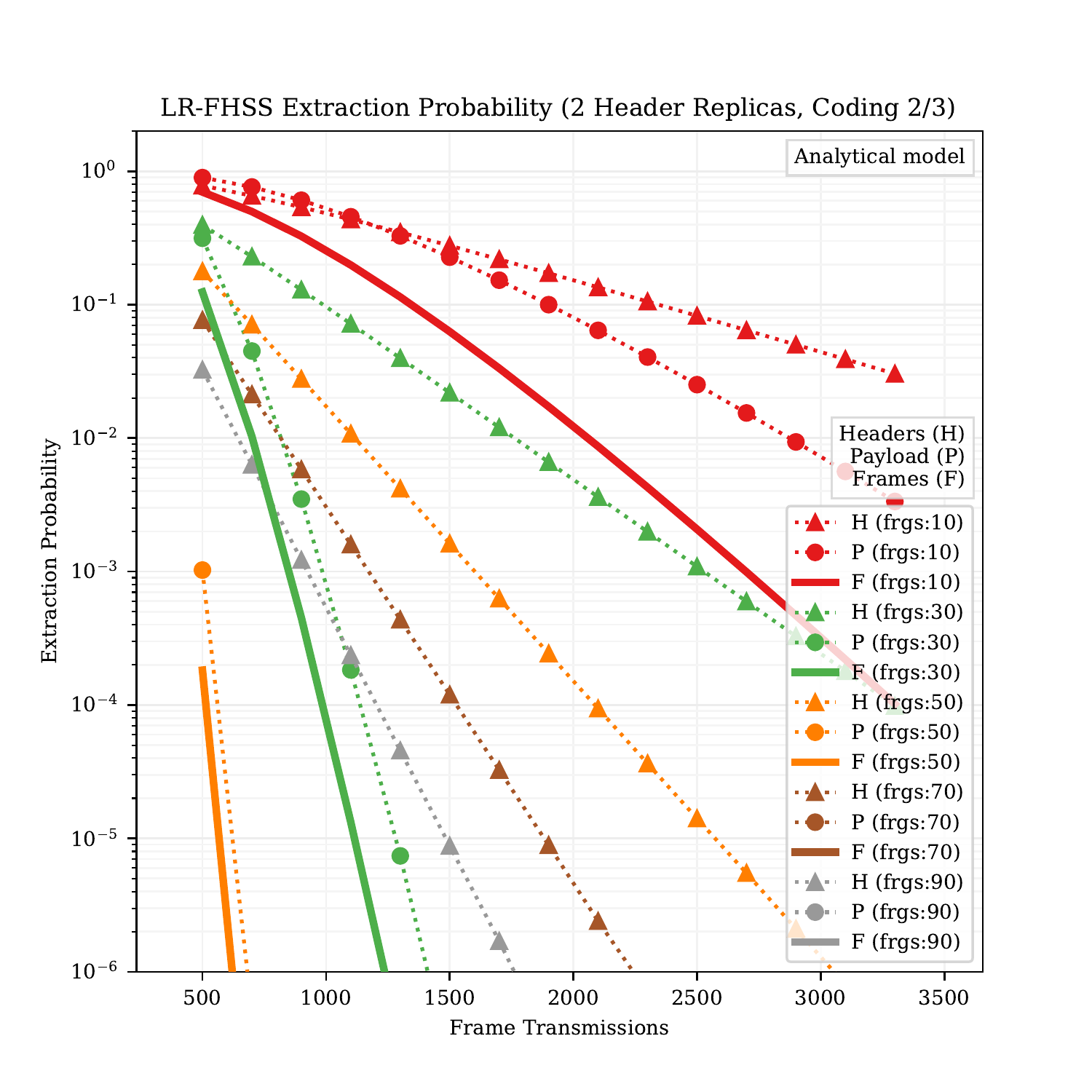}}
    \subfloat[]{\includegraphics[width=.49\textwidth,trim={0.6cm 0.6cm 1cm 0cm},clip]{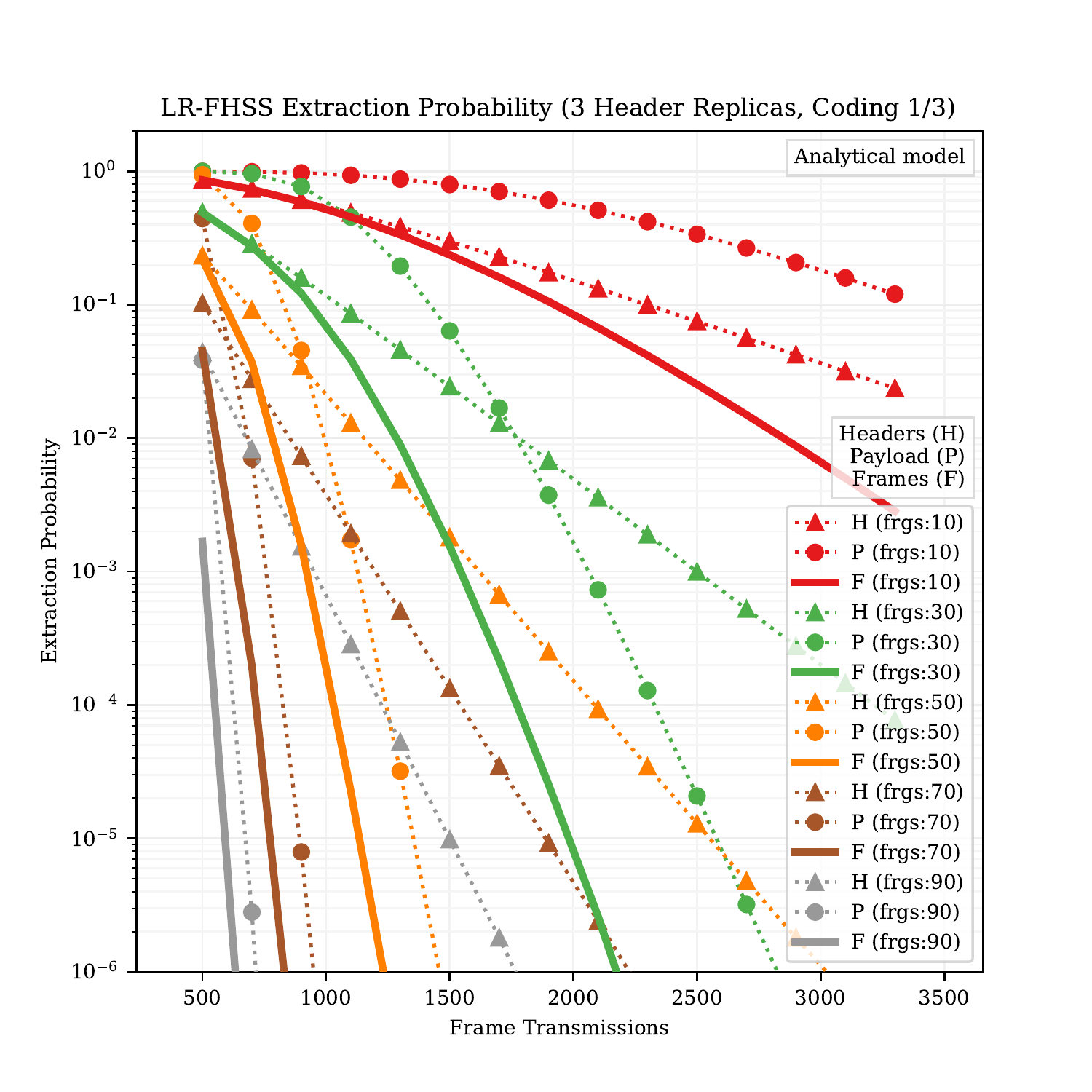}}
    \caption{Analytical extraction model. Successfully extracted headers (H), payload fragments (P), and complete frames (F) for varying frame transmission count for the more aggressive configuration (a) and the most robust LR-FHSS configuration (b).}
    \label{fig:extraction}
\end{figure*}

We denote $p_{hdr}$ the probability of receiving at least one header replica.
If $coll(233)$ is the probability that a collision occurs during the 233~ms of any of the three replicas sent in the robust mode, then $$p_{hdr}=1-coll(233)^3.$$

On the other hand, the payload is received if at least one-third of the fragments are received in the robust configuration. We denote $p_{pld}$ to the probability of receiving the required number of fragments. Therefore, we have:
$$p_{pld}=\sum_{i=\lceil{}\mathcal{P}/3\rceil{}}^{\mathcal{P}}{\mathcal{P} \choose i}.(1-coll(102))^i.coll(102)^{\mathcal{P}-i},$$
where $\mathcal{P}$ is the number of fragments and $coll(102)$ denotes the probability that a collision occurs during the 102~ms of each fragment. Note that the term within the sum refers to the probability that exactly $i$ fragments out of $\mathcal{P}$ are received.

Finally, an \lr frame is received if both the header and the payload are received. The probability of receiving a complete frame is thus:
$$p_{frame}=p_{hdr}.p_{pld}.$$

The probability of collision $coll(d)$ on a given OBW during a period $d$ depends on the number of ongoing transmissions $n_{og}$ at a given time slot. Let us assume for simplicity that all transmissions are synchronized at the slot level, where each slot spans 102~ms. If we denote by $\mathcal{C}$ the number of OBWs, the average number of free OBWs at a given time slot is equal to:
$$free(\mathcal{C})=\mathcal{C}.(1-1/\mathcal{C})^{n_{og}}.$$
Thus, the probability that a collision occurs during a duration $d$ is:
$$coll(d)=1-(free(\mathcal{C})/\mathcal{C})^{\lceil{}d/102\rceil}.$$
Note that if there are exactly $n_{tx}$ full transmissions during a number $\mathcal{T}$ of 102~ms time slots, then we have:
$$n_{og}\approx{}(3*233+\mathcal{P}*102)n_{tx}/(102\cdot\mathcal{T}).$$
% ===> https://math.stackexchange.com/questions/2833309/expected-number-of-empty-boxes-placing-m-balls-into-n-boxes

Figure~\ref{fig:extraction} shows the probability of header, payload, and full frame reception for the fast configuration (Figure~\ref{fig:extraction}(a)) and for the robust configuration (Figure~\ref{fig:extraction}(b)). The probabilities are given as a function of the number of concurrent transmissions $n_{tx}$, during $\mathcal{T}=1000$ time slots, and for a varying number of fragments $\mathcal{P}\in{}\{10,30,50,70,90\}$. In all cases, the probability of losing the header plays a significant role in the overall frame loss. This is particularly true in the robust configuration, where the header extraction probability can be lower than the payload recovery probability. 
For instance, for 1000 frame transmissions and a frame length of 30 fragments, headers are lost for about 10\% of the frames in both configurations. However, the payload is lost for 99.9\% of the frames in the fast configuration and for 60\% of the frames in the robust configuration. 
The observed relative fragility of \lr headers is consistent with the results from~\cite{ullah22analysis} for DR8\footnote{see Figure 6 and the conclusion of~\cite{ullah22analysis}.}, and from~\cite{maleki2022d2d}\footnote{see Figure 8 of~\cite{maleki2022d2d}.}.

\section{Headerless Frames Recovery Methods}
\label{section:proposal}

To address the problem of recovering headerless LR-FHSS transmissions, this section describes two solutions. The first uses an ILP model, and the second is a cost-effective heuristic. We start by describing our assumptions and framing a model on which to base these methods.

\subsection{Assumptions}

% Assumptions
{\bf Channel:} We focus on a single OCW and a single grid to simplify the approach. Extending the solutions to multiple OCWs and grids is trivial (\eg, one model instance per OCW/grid). Thanks to the robust GMSK modulation used for fragments, we assume that fragments always reach the gateway (there is no loss due to weak signal) and there is no noise in the channel: in other words, an OBW is always detected free when there is no transmission using it, and always detected busy if there is at least one transmission using it. 
We consider that fragments collide if they overlap in time and use the same OBW, that is, we consider the pessimistic case where there is no capture effect. Fragment collisions can be detected using channel sensing, but they can never be resolved with our assumption.

{\bf Frames:} We assume that the number $\mathcal{P}$ of fragments per frame is known and fixed, that is, all frames have the same length. We also assume that all transmissions start and end within the $\mathcal{T}$ time slots of the whole simulation, which is a weak assumption when $\mathcal{T}$ is much larger than the duration of a single frame. Finally, we assume a time-slotted model where each slot lasts for the duration of one fragment, and each transmission starts at the beginning of a slot.

We discuss the effect of these assumptions on the proposed solutions in Subsection~\ref{subsection:discussion-assumptions}. We can already mention here that some of them (\eg, no capture effect) correspond to a worst-case scenario, and are therefore pessimistic. By relaxing such assumptions, the performance of our algorithms would improve, but this requires obtaining real evaluation data on the \lr modulation.

\subsection{Problem model}

\begin{figure}[htbp]
    \centering
    \includegraphics[width=0.75\linewidth]{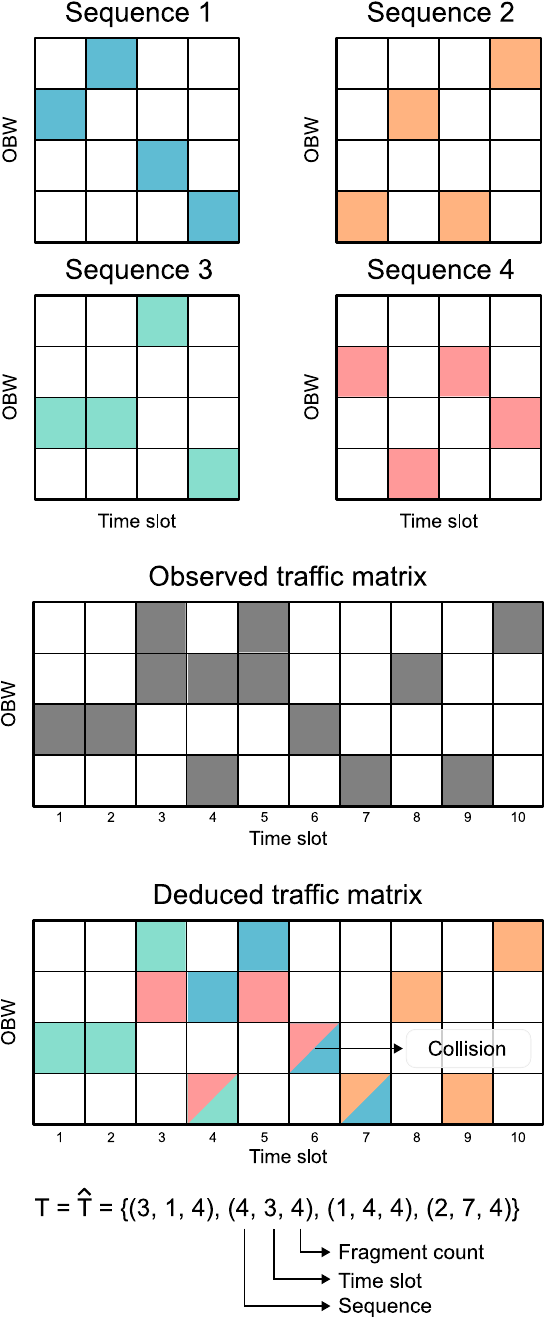}
    \caption{A simplified example of a grid with 4 OBWs and 4 frequency hopping sequences of the same length. Each sequence is represented by a different color. The observed traffic matrix $M[t][c]$ represents the channel activity detected by the gateway: each square corresponds to an $M[t][c]$ element and is colored gray if $M[t][c]=1$, meaning an activity (one or more fragment transmissions) has been observed on OBW $c$ at time $t$. The deduced traffic matrix $\hat{T}=\{(s_i,t_i, p_i)\}$) contains all the sequences that have been identified from the observed traffic matrix $M$. As we can see in the example each square that was gray in the observed traffic matrix $M$ has been changed in the color of the identified sequence. If a square has more than one color, it means two fragments from different frames (and hence different sequences) collided. If all the gray squares can be colored in a given sequence it means, that all frames have been correctly identified, and hence $T=\hat{T}$. Our objective in this paper is to propose a solution that maximizes the number of identified sequences (and hence frames).}
    \label{fig:sequences}
\end{figure}

Our model is comprised of the following elements summarized in Table~\ref{tab:model} and described below.

% Constants
We define a series of constants to model the LR-FHSS environment. We define the number of OBWs per grid $\mathcal{C}$, the number of time slots $\mathcal{T}$, the number of possible pseudo-random sequences $\mathcal{S}$, the number of transmitted frames $\mathcal{F}$, and the number of fragments used on each frame $\mathcal{P}$.

\begin{table}[]
\centering
\caption{LR-FHSS Constants and Variables.}
\label{tab:model}
\begin{tabular}{|ll|}
\hline
\multicolumn{2}{|c|}{Constants}            \\ \hline
% \multicolumn{1}{|c|}{Symbol} & Description \\ \hline
\multicolumn{1}{|c|}{$\mathcal{C}$} &  Number of OBWs per grid \\ \hline
\multicolumn{1}{|c|}{$\mathcal{T}$} &  Number of time slots \\ \hline
\multicolumn{1}{|c|}{$\mathcal{S}$} &  Number of sequences \\ \hline
\multicolumn{1}{|c|}{$\mathcal{F}$} &  Number of frames \\ \hline
\multicolumn{1}{|c|}{$\mathcal{P}$} &  Number of fragments per frame \\ \hline \hline
\multicolumn{2}{|c|}{Variables}            \\ \hline
\multicolumn{1}{|c|}{$S = \{s \mid s \in \mathcal{S} \}$} & \begin{tabular}[c]{@{}l@{}}Set of $\mathcal{S}$ sequences\end{tabular} \\ \hline
\multicolumn{1}{|c|}{$M[t][c] = \{0,1\}$} & \begin{tabular}[c]{@{}l@{}}Matrix ($\mathcal{T} \times \mathcal{C}$) of observed\\transmissions equal to $1$ if one or more\\transmissions were observed on OBW\\$c \in [1;\mathcal{C}]$ at time $t \in [1;\mathcal{T}]$ \end{tabular} \\ \hline
\multicolumn{1}{|c|}{$T=\{(s_i,t_i, p_i)\}$} & \begin{tabular}[c]{@{}l@{}}List of $\mathcal{F}$ 3-tuples of frame\\transmissions each defined by a\\sequence $s_i\in{}S$, a time $t_i \in [1;\mathcal{T}]$ and its \\length (in number of fragments) $p_i=\mathcal{P}$\end{tabular} \\ \hline
\multicolumn{1}{|c|}{$\hat{T}$} & \begin{tabular}[c]{@{}l@{}}The frame transmission list (same\\structure as $T$) that the gateway was\\able to derive from observed fragment\\receptions in $M[t][c]$\end{tabular} \\ \hline
\multicolumn{1}{|c|}{$y_{t,s}\in{}\{0,1\}$} & \begin{tabular}[c]{@{}l@{}}A matrix ($\mathcal{T}\times\mathcal{S}$) representation of $\hat{T}$ \end{tabular} \\ \hline
\end{tabular}
\end{table}

% Variables
We also define the following variables. $S$ is a set of $\mathcal{S}$ random sequences, each representing the frequency hopping pattern (see the example of sequences from Fig.~\ref{fig:sequences}).
Furthermore, we define a two-dimensional boolean matrix $M$ called the observed traffic matrix, such that $M[t][c]=1$ if and only if one or more fragment transmissions were observed on OBW $c$ at time $t$.
A matrix $M$ contains all OBW in a given grid.
We assume no fragment losses due to weak signal reception, but we consider the possibility of collisions. 
Therefore, the matrix $M$ is composed of a series of fragment transmissions for each frame using a given sequence from $\mathcal{S}$.
As a result, each frame transmission $i$ is uniquely defined by a sequence $s_i \in S$, a starting time slot $t_i \in [1;\mathcal{T}]$, and its length $p_i=\mathcal{P}$.
We coin $T$ as the set of 3-tuple elements $\{(s_i, t_i, p_i)\}$, each 3-tuple defining a single frame transmission.

% Objective
While $T$ is the actual transmission set, the receiving gateway can only observe the matrix $M$.
Indeed, $M$ comprises all fragment transmissions without the underlying frame structure (see the observed traffic matrix in Fig.~\ref{fig:sequences}).
In fact, the core problem of decoding headerless \lr transmissions is to correctly map observed fragment receptions from $M$ to the actual frame transmissions of $T$, including the sequence, time, and length $(s_i, t_i, p_i)$ for each frame $i$ (see the deduced traffic matrix in Fig.~\ref{fig:sequences}).
The goal is for the gateway to derive a frame transmission set $\hat{T}$ as close as possible to the original $T$.
In the ideal case, all transmitted frames can be recovered when $T = \hat{T}$.

\subsection{ILP Model}

% MILP Model
Based on the constants and variables from Table~\ref{tab:model}, the headerless decoding problem for the \lr modulation can be formalized as the following optimization problem.

\begin{equation*}
\begin{array}{ll@{}ll}
\text{min.}  & \displaystyle\sum\limits_{t=1}^{\mathcal{T}} \sum\limits_{s=1}^{\mathcal{S}} y_{t,s} & &
\\ \\
\text{s.t.}& \displaystyle \mathcal{P} \cdot y_{t,s} \geq \displaystyle\sum\limits_{k=1}^{\mathcal{P}} M_{t+k-1,s[k]}  & & \forall{}t\in{}[1;\mathcal{T}], \forall{}s\in{}S.
\end{array}
\end{equation*}

For ease of manipulation by the ILP, we have introduced a new set of variables $y_{t,s}$, $\forall{}t\in[1;\mathcal{T}]$ and $\forall{}s\in{}S$, such that $y_{t,s}=1$ if and only if $(s,t,\mathcal{P})\in{}\hat{T}$.
The resulting model aims at finding the minimum number of transmissions in $\hat{T}$ needed to cover the matrix $M$.
To this end, the objective function aims to minimize the number of sequences in $\hat{T}$ ($y_{t,s}$) for all time slots $t$ and sequences $s$.
However, the set of transmissions in $\hat{T}$ must comply with a constraint that forces each frame's $\mathcal{P}$ fragments to correspond to an observed OBW occupancy of 1 in $M$.
As observed in the previous equations, a variable number of fragments would render a non-linear model. The linearization of this non-linear model is left as future work.
To solve the theoretical model with state-of-the-art linear solvers, we assume $\mathcal{P}$ is constant for all transmissions in the subsequent evaluation.

\subsection{Heuristic Solution}

Besides the previous exact model, we introduce a simple yet effective heuristic to find $\hat{T}$ from the input matrix $M$. The step-by-step process is listed in Algorithm~\ref{alg:heuristic}. 
The heuristic iterates over each time slot $t$ in an increasing manner. Then, for each sequence $s$ among the possible set of 512 known sequences, it greedily determines whether $M$ contains a transmission starting at this time slot $t$ and with sequence $s$. If this is the case, $(s,t,\mathcal{P})$ is added to $\hat{T}$. This greedy heuristic has a time complexity of $\mathcal{O}(\mathcal{T}\cdot\mathcal{S}\cdot\mathcal{P})$ when the full matrix $M$ is known statically. Indeed, the main operation of line 6 is performed within three nested {\tt for} loops, the first (line 2) being executed $\mathcal{T}$ times, the second (line 3) being executed $\mathcal{S}$ times, and the third (line 5) being executed $\mathcal{P}$ times.

Since $[1;\mathcal{T}] increasing time slots explore $, this static heuristic can be easily translated into an online heuristic with a sliding window of $\mathcal{P}$ time slots, thus resulting in a time and space complexity of $\mathcal{O}(\mathcal{S}\cdot\mathcal{P})$ at each time slot. Note that according to the specification of \lr, we have $\mathcal{S}=2^9$ and $\mathcal{P}\leq{}32$. The sliding-window capability, combined with the reduced complexity, makes this heuristic a perfect candidate to be implemented in future \lr gateways onboard resource-constrained LEO satellites to enable the recovery of \lr frames in which headers are lost.

\begin{algorithm}
\caption{Greedy Headerless Decoding Heuristic.}
\label{alg:heuristic}
\begin{algorithmic}[1]
\Require $\mathcal{T}$, $S$, $\mathcal{P}$ and $M$
\State $\hat{T} \gets \{ \}$
\For{$t \in [1;\mathcal{T}]$}
    \For{$s \in S$}
        \State $found \gets true$
        \For{$p \in [1;\mathcal{P}]$}
            \If{$M[t + p-1][s[p-1]] \neq 1$}
                \State $found \gets false$
            \EndIf
        \EndFor
        \If{$found=true$}
            \State $\hat{T} \gets \hat{T}\cup\{ (s, t, \mathcal{P}) \}$
        \EndIf
    \EndFor
    \State return $\hat{T}$
\EndFor
\end{algorithmic}
\end{algorithm}

\section{Simulation results}
\label{section:simulation-results}

\begin{table*}[]
\centering
\caption{Simulation Parameters.}
\begin{tabular}{|l|l|}
\hline
Description                   & Value(s)                    \\ \hline \hline
Number of OCW                 & 1                           \\ \hline
Number of grids               & 1                           \\ \hline
Number of OBWs per grid       & 35                          \\ \hline
Number of time slots $\mathcal{T}$ & 1000                        \\ \hline
Number of sequences $\mathcal{S}$ & 512                         \\ \hline
Number of frames $\mathcal{F}$ & 500 to 3200 in steps of 100 \\ \hline
Number of fragments $\mathcal{P}$ per frame & 10 to 90 in steps of 20     \\ \hline
Runs per step                 & 10                          \\ \hline
\end{tabular}
\label{table:params}
\end{table*}

\begin{figure*}[htbp]
    \centering
    % % trim={<left> <lower> <right> <upper>}
    \subfloat[]{\includegraphics[width=.49\textwidth,trim={0.6cm 0.6cm 1cm 1cm},clip]{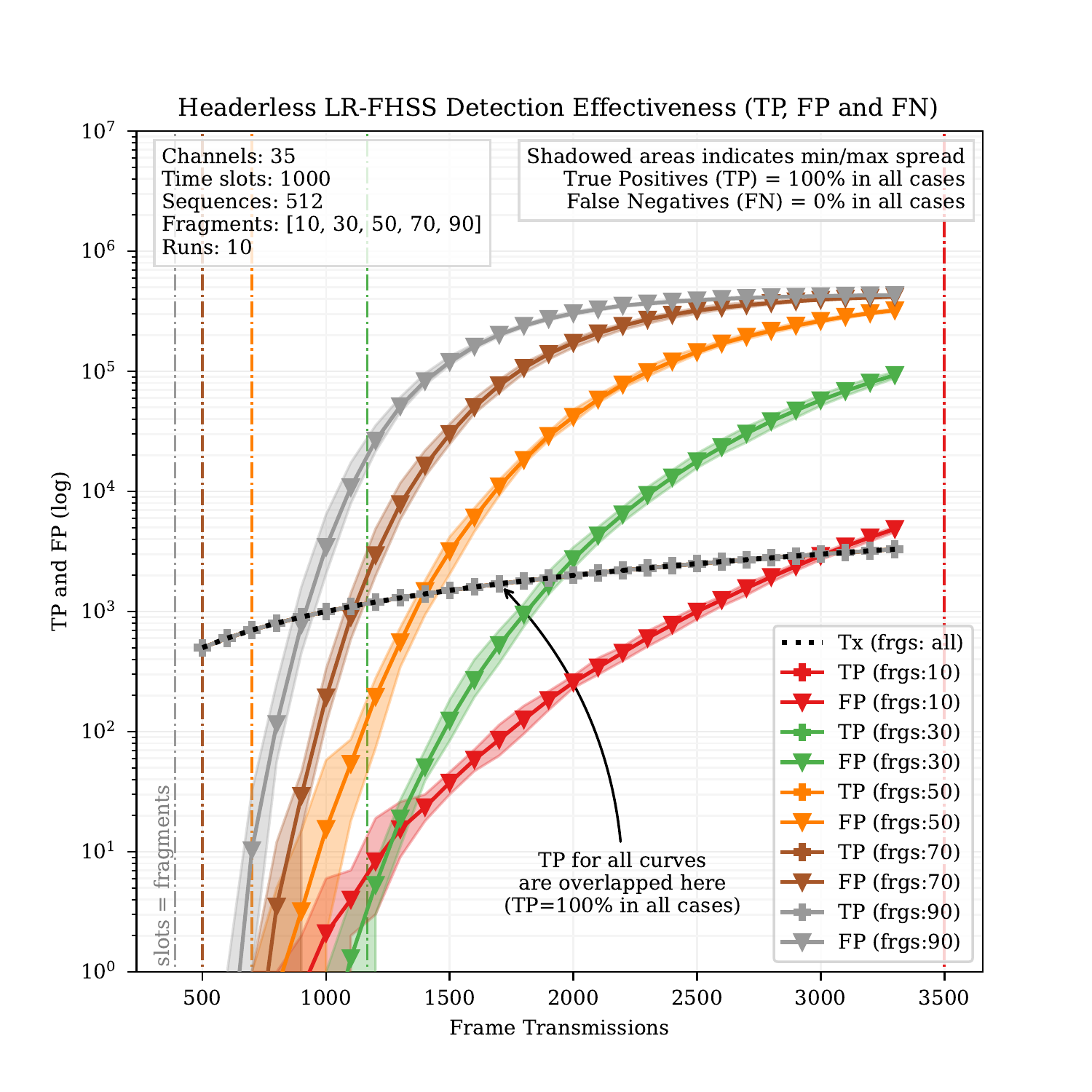}}
    \subfloat[]{\includegraphics[width=.49\textwidth,trim={0.6cm 0.6cm 1cm 1cm},clip]{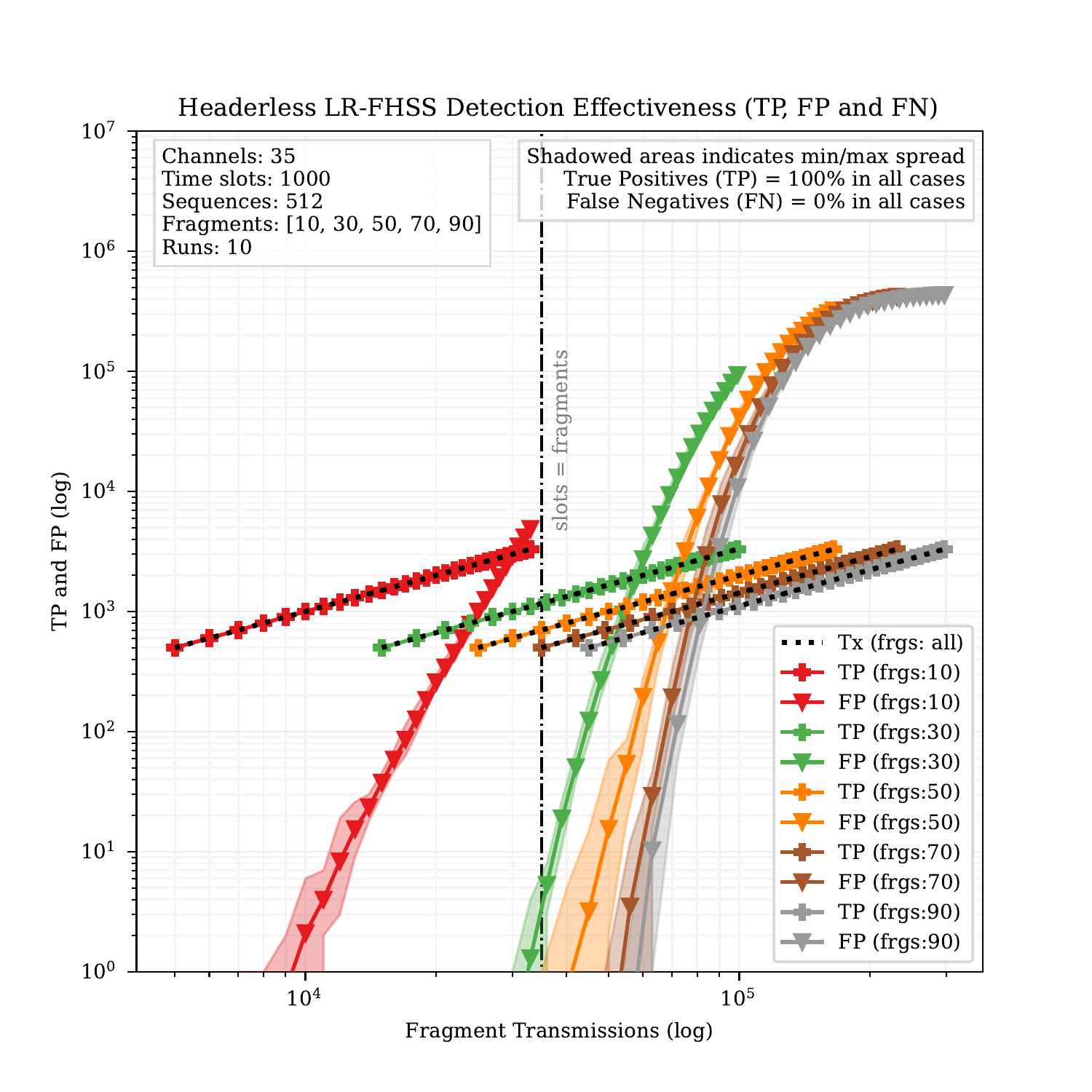}}
    \\
    \subfloat[]{\includegraphics[width=.49\textwidth,trim={0.6cm 0cm 1cm 0.4cm},clip]{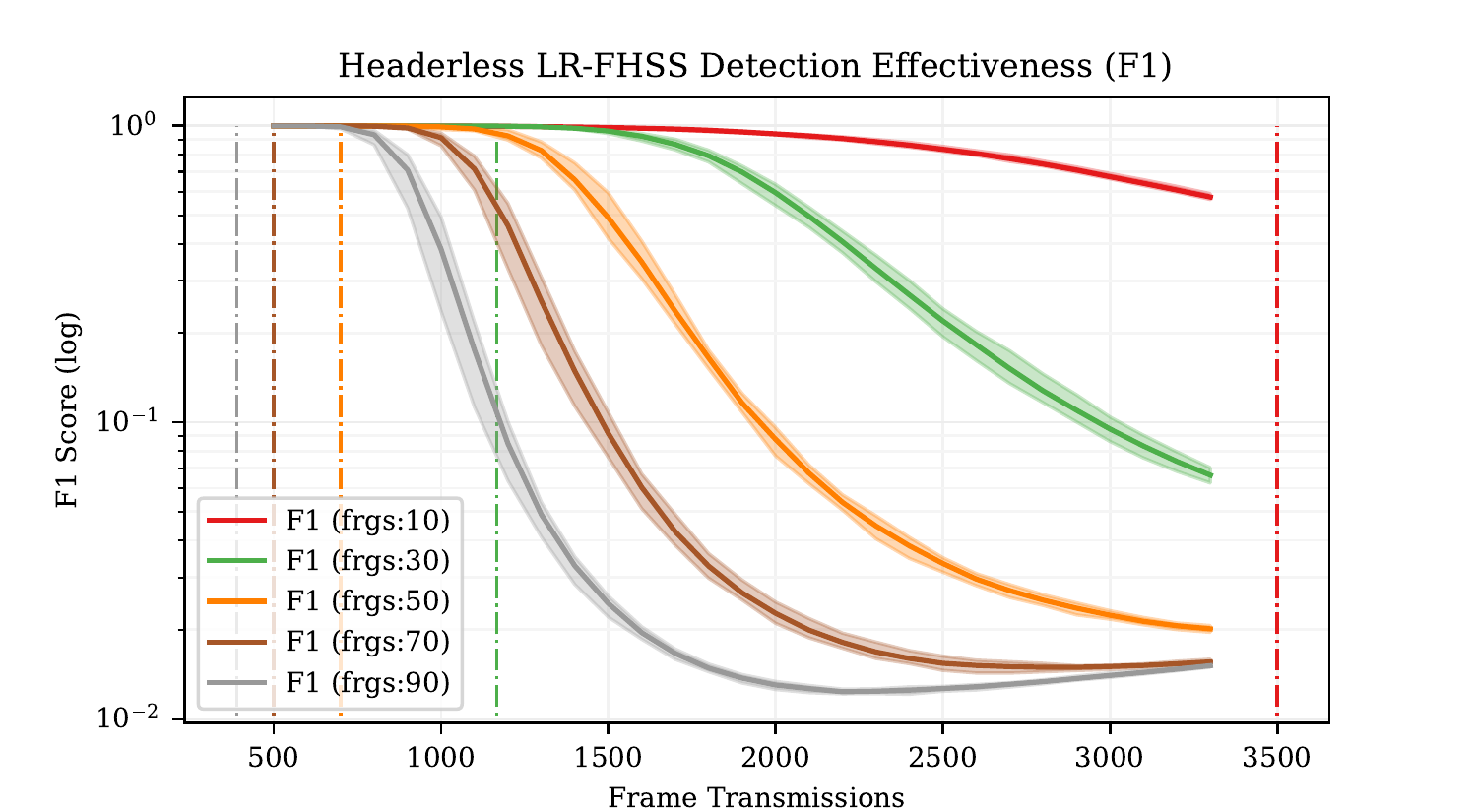}}
    \subfloat[]{\includegraphics[width=.49\textwidth,trim={0.6cm 0cm 1cm 0.4cm},clip]{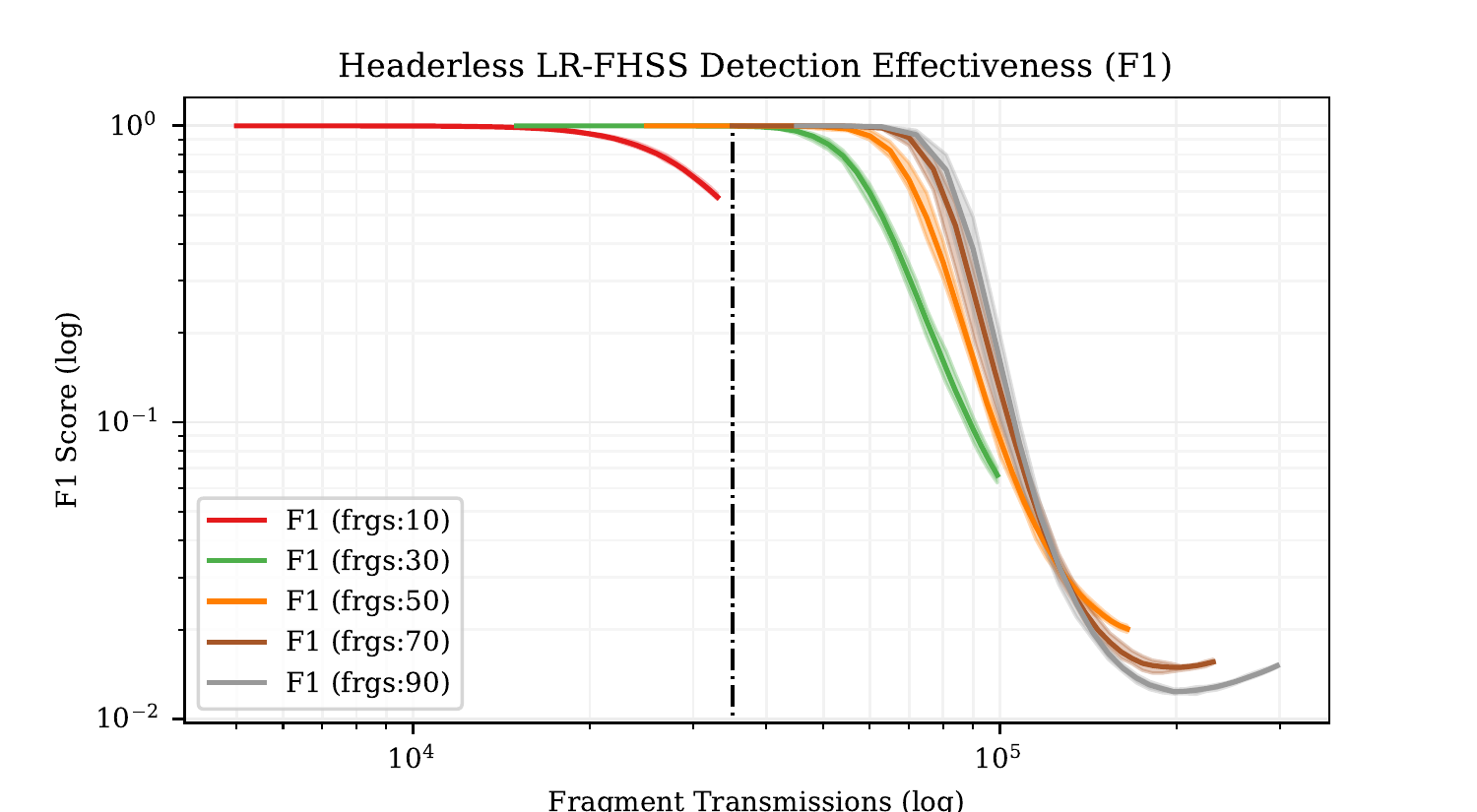}}
    \\
    \subfloat[]{\includegraphics[width=.49\textwidth,trim={0.6cm 0cm 1cm 0.4cm},clip]{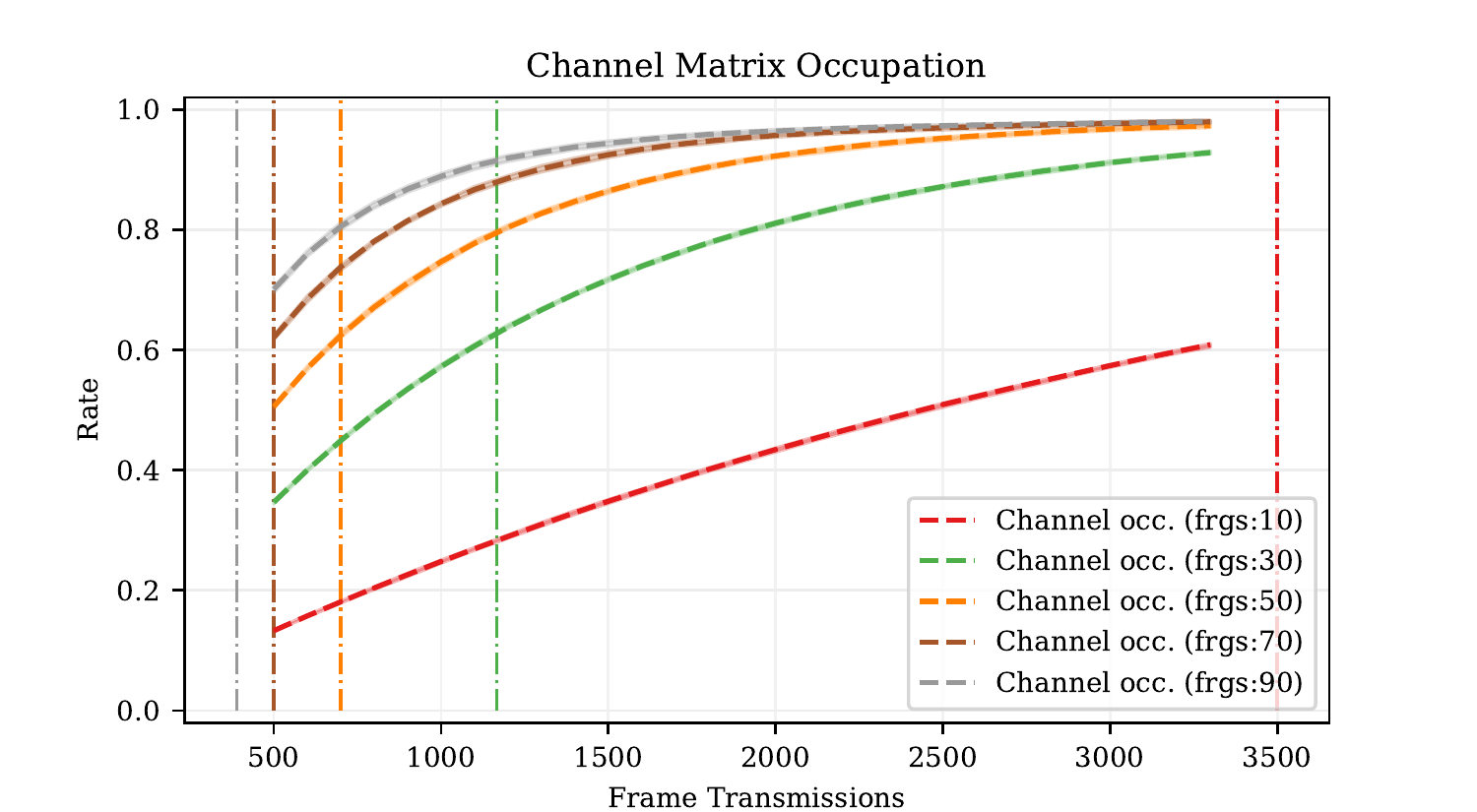}}
    \subfloat[]{\includegraphics[width=.49\textwidth,trim={0.6cm 0cm 1cm 0.4cm},clip]{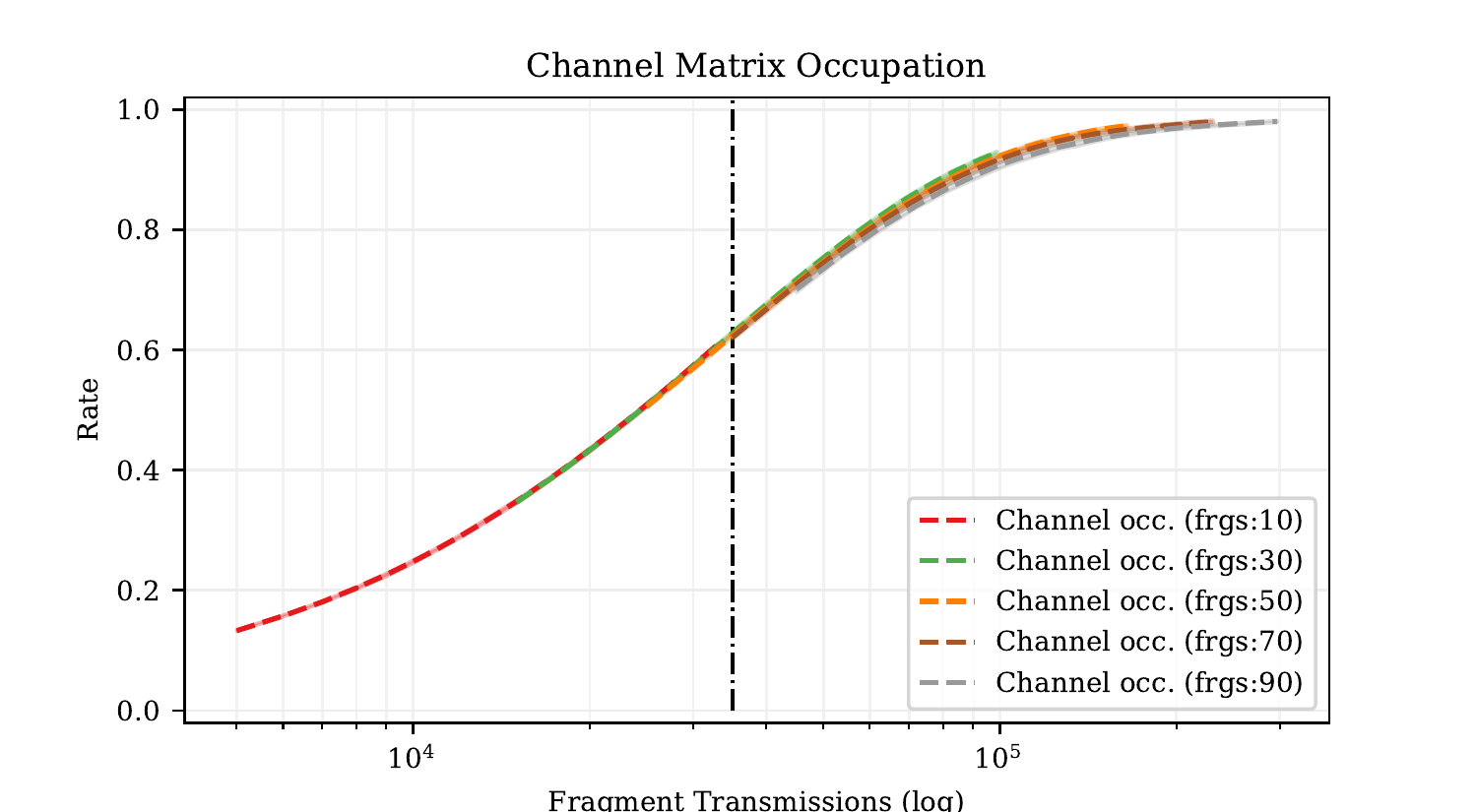}}
    \caption{Detection effectiveness results. True Positives (TP), False Positives (FP), and False Negatives (FN) for varying fragment count expressed vs. (a) frame transmissions and (b) total fragment count. The F1 score and OBW matrix occupation are shown in (c), (d), (e), and (f) for frame transmissions and total fragment count, respectively. Plots are valid for the ILP model and the heuristic algorithm as they provide exactly the same effectiveness results in all cases.}
    \label{fig:tpfp}
\end{figure*}

% Simulation setup
We conducted an extensive simulation campaign to evaluate the effectiveness and efficiency of the ILP model and the heuristic approach.
We implemented a reference Python library \footnote{Link to repository omitted in blind-review.} that leverages an interface with Gurobi~\cite{gurobi} to solve the ILP model.
The parameters used in the following analysis are summarized in Table~\ref{table:params}.
% new phrase
Since the channel hopping of LR-FHSS takes place on a single grid, we considered here one grid and one OCW.
The performance results can be easily generalized for LR-FHSS systems with more grids or more OCWs. We used the European settings for the number of OBWs per grid (35) and the number of sequences (512).
We randomly generate a new set of sequences $S$ and a new list of frame transmissions $\mathcal{T}$ (uniform distribution) for each simulation run. 
For each metric, we plot the average with lines and the minimum and maximum values with a shaded area.
We allow repeated transmission elements in $T$ (same $(s,t,p)$ 3-tuple can occur multiple times), but not in $S$.

% Effectiveness
\subsection{Detection Effectiveness} 
\label{sec:detection}

% Metrics
The metrics to assess the decoding success are based on the difference between the real traffic $T$ and the inferred traffic $\hat{T}$. 
Recalling that both traffic matrices are comprised by $(s_i,t_i,p_i)$ tuples where $t_i$ is the time at which a frame starts with sequence $s_i$ and with $p_i=\mathcal{P}$ fragments, the effectiveness metrics are:
\begin{itemize}
    \item True positive (TP): $(s,t)\in{}T$ and $(s,t)\in{}\hat{T}$,
    \item False positive (FP): $(s,t)\notin{}T$ but $(s,t)\in{}\hat{T}$,
    \item False negative (FN): $(s,t)\in{}T$ but $(s,t)\notin{}\hat{T}$.
    \item F1-Score: $F1 = 2 \times TP / (2 \times TP + FP +FN)$.
\end{itemize}
The F1 Score is the harmonic mean of \textit{precision} (accuracy of the model in identifying positive instances) and \textit{recall} (completeness of the model in identifying positive instances), combining both metrics into a single value.
Thus, high decoding effectiveness is achieved by high TP and low FP and FN. Also, a high F1 score indicates better performance and accuracy.

% Control variables
We analyze these metrics for varying traffic loads. Traffic load can be expressed in two ways: by the number of transmitted frames or the total fragment count.
Frame decoding rate is a better indicator of the user experience, while fragment count is a direct measure of the OBW occupation of matrix $M$. Thus, we will use both traffic loads in the following results.

% Figure overview
Figure~\ref{fig:tpfp} shows the effectiveness (TP, FP, and FN), of both the ILP and the heuristic, as a function of the traffic load. 
Figure~\ref{fig:tpfp}~(a) uses frame transmission as traffic load, and Figure~\ref{fig:tpfp} uses fragment count (in logarithmic scale). 
Figure~\ref{fig:tpfp}~(c) and~(d) offer F1, a more synthetic metric for all numbers of fragments.
In these figures, to indicate the channel occupation, the vertical lines represent a channel occupancy such that the number of used slots in $M$ equals the number of fragments among all transmissions in $T$. 
As a more precise channel utilization reference, Figure~\ref{fig:tpfp}~(e) and~(f) present the channel matrix occupation measured in $M$. This last curve confirms that plots expressed against fragment transmissions are compared over similar channel conditions.

%  Control variables specifics
Results are obtained from simulations of 500 to 3300 frame transmissions (with steps of 100) for 10, 30, 50, 70, and 90 fragments each (total fragment count ranges from 5000 to 297000).
The time horizon is 1000 time slots, and we assume 512 sequences over 35 OBWs as specified for LR-FHSS~\cite{lorawan-rp2103}.
The resulting configuration corresponds to robust (DR8) and fast (DR9) data rates from Table~\ref{tab:reg_params} (coding rate and bit rate will become relevant in the extraction efficiency discussed in Section~\ref{sec:extraction}).
For each step, we perform 10 simulations (random $s$ and $t$ for each run) and record the mean, maximum, and minimum for each metric (shadowed area in Figure~\ref{fig:tpfp}).

% Observation 1: model same as algo
The first observation from the simulation results is that the ILP model and the heuristic algorithm provide precisely the same TP, FP, and FN results.
This is a promising outcome considering the algorithm can operate without relying on an ILP solver while delivering optimal performance in the aforementioned settings.

% Observation 2: Perfect TP
The second observation is that the $(s_i,t_i,p_i)$ tuples from $T$ are always contained in $\hat{T}$. 
In other words, TP is always 100 \%, which indicates that all transmitted headerless frames can be detected at the receiver end in all evaluated cases.
However, FP increases exponentially with the traffic load to the point that FP exceeds the number of TP.
Consequently, the gateway will detect many incorrect frames when the channel is heavily loaded.
Obtained F1 scores support this statement (Figure~\ref{fig:tpfp}~(c) and~(d)), as optimum precision/recall trade-off is provided until the channel matrix becomes heavily occupied.
However, as discussed in Section~\ref{section:discussion}, the CRC checks of LoRaWAN can detect and filter most of these FP frames.

\begin{figure*}[htbp]
    \centering
    \subfloat[]{\includegraphics[width=.49\textwidth,trim={0.58cm 0.6cm 1cm 1cm},clip]{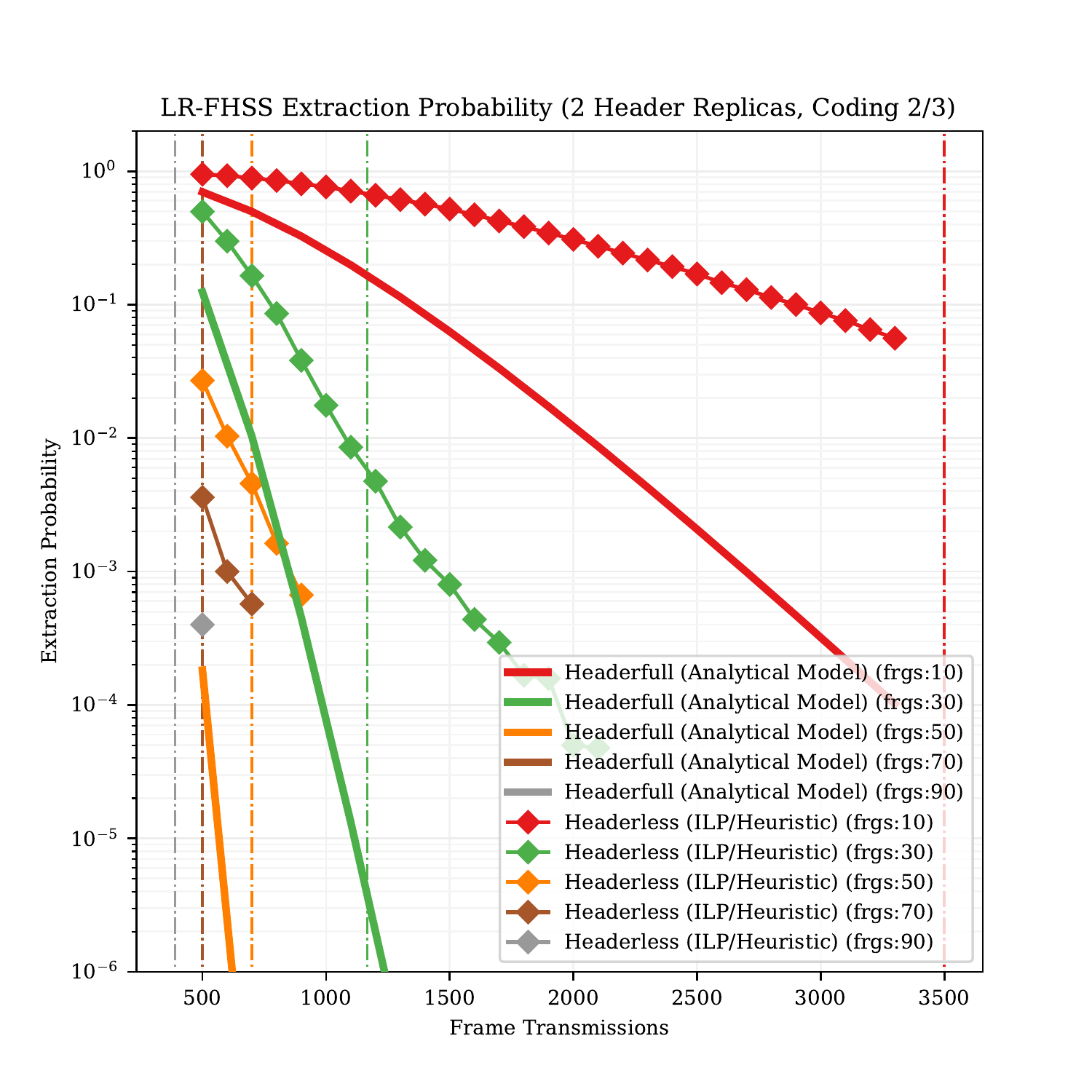}}
    \subfloat[]{\includegraphics[width=.49\textwidth,trim={0.58cm 0.6cm 1cm 1cm},clip]{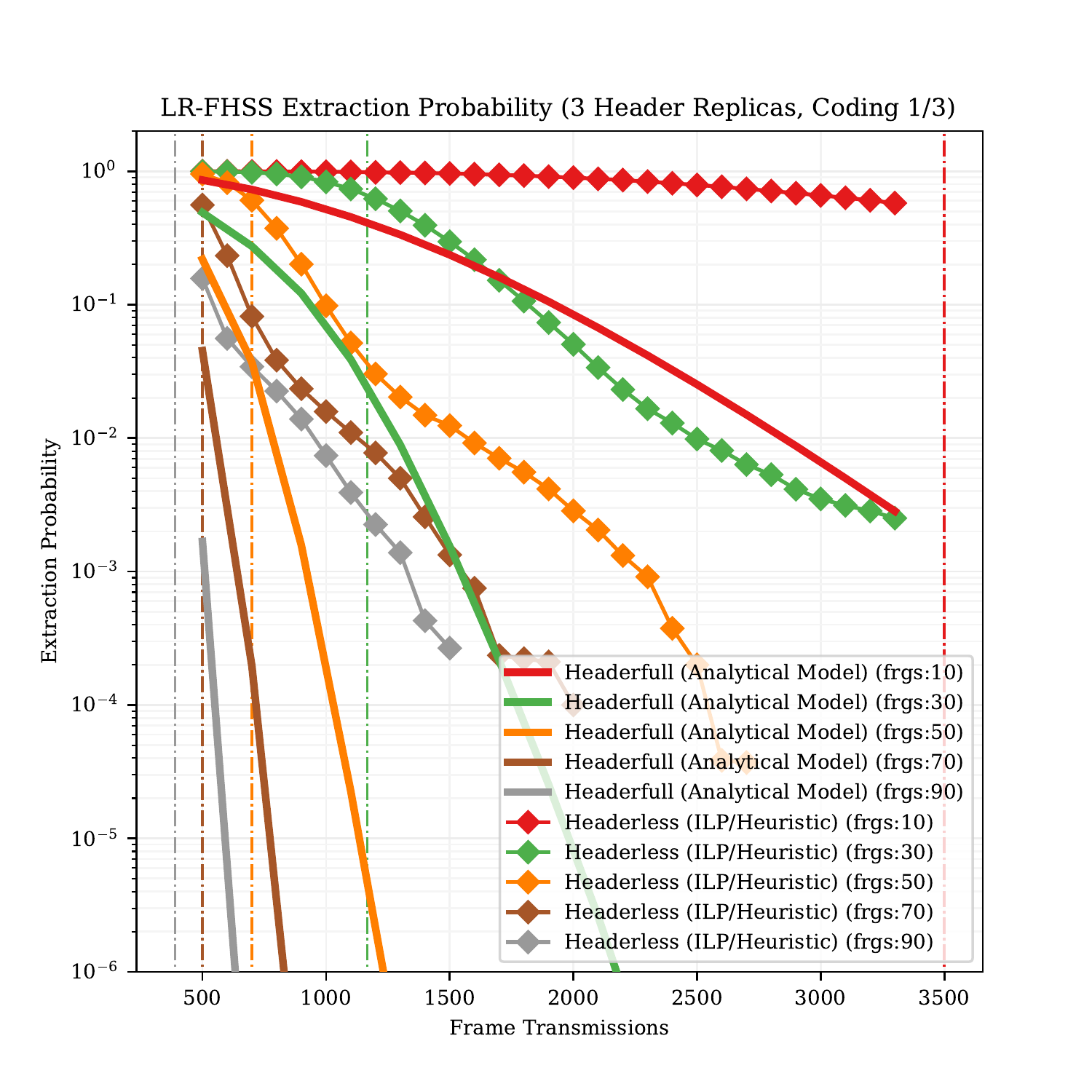}}
    \caption{Extraction effectiveness results. Plots compare the headerfull extraction rate model from Section~\ref{section:motivation} with the headerless extraction rate using the ILP and the heuristic approach presented in Section~\ref{section:proposal}. Results are for (a) the fast configuration and (b) the robust configuration.}
    \label{fig:fig5}
\end{figure*}

\begin{figure*}[htbp]
    \centering
    % trim={<left> <lower> <right> <upper>}
    \subfloat[]{\includegraphics[width=.49\textwidth,trim={0.6cm 0.6cm 1cm 1cm},clip]{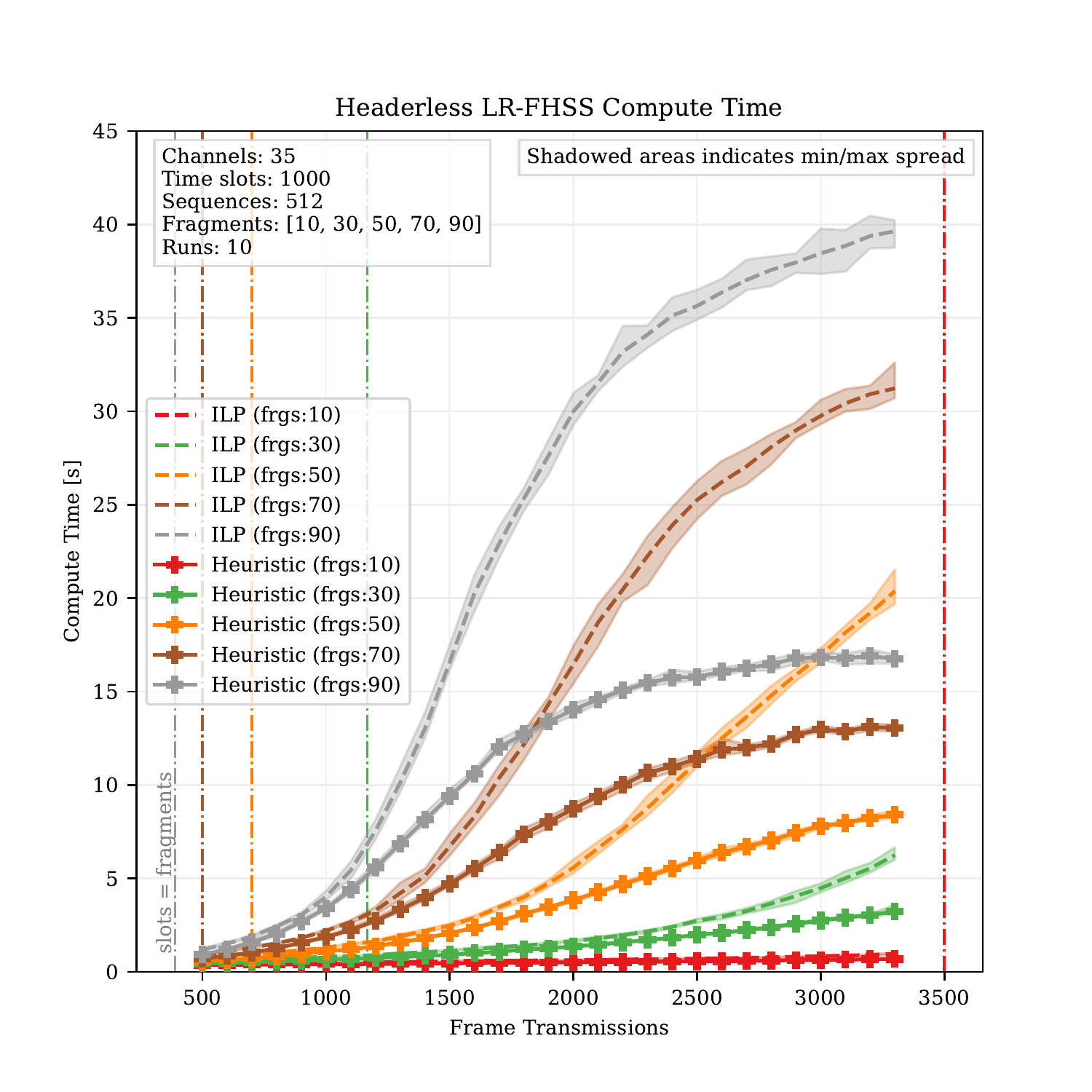}}
    \subfloat[]{\includegraphics[width=.49\textwidth,trim={0.6cm 0.6cm 1cm 1cm},clip]{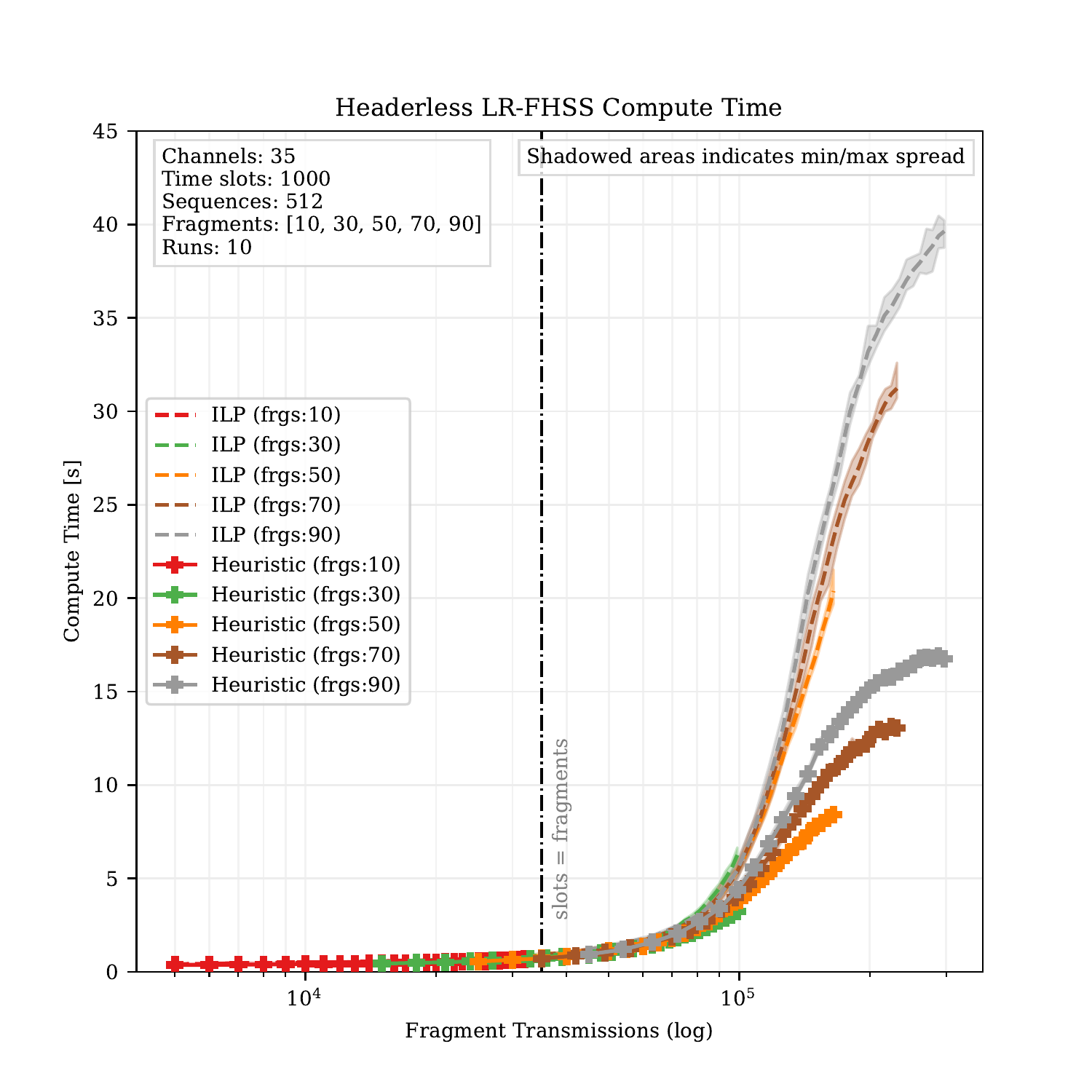}}
    \caption{Computation efficiency results. Compute time for the ILP model and the heuristic for varying fragment count expressed vs. (a) frame transmissions and (b) total fragment count.}
    \label{fig:ctime}
\end{figure*}

% Observation 3: fragment
The third observation involves the impact of fragment count.
At first glance, results show that transmissions with smaller frames (that is, fewer fragments) provide higher efficiency thanks to a reduced FP count (Figure~\ref{fig:tpfp}~(a)).
However, this is a direct consequence of a less occupied channel, as frames with fewer fragments require fewer slots in $M$.
If we observe the FP measured in fragment transmissions (Figure~\ref{fig:tpfp}~(b)), we find that for equal OBW occupation, larger frames are more effective.
Indeed, larger frames exploit more extensive sequences $s$ and have better chances of being correctly identified at the receiver.
This finding motivates the utilization of frame aggregation features in the device-to-gateway link.

% Channel occupation
A better appraisal of OBW occupation can be achieved by observing the vertical lines in both Figures~\ref{fig:tpfp}(a) and (b).
These lines indicate the point at which there are as many transmitted fragments as total slots in $M$.
Namely, if no collisions exist, $M$ would be fully occupied at these lines. But due to collisions, $M$ is filled at approximately 60\% at this line.
Of course, in Figure~\ref{fig:tpfp}~(a), the x-position depends on the number of fragments per frame. 
But, in Figure~\ref{fig:tpfp}~(b), the maximum fragment count to fill the channel is equal to 35000 ($\mathcal{C}\times\mathcal{T}=35\times{}1000=35000$). Figures~\ref{fig:tpfp}~(c) and~(d) show the occupancy of the channel matrix (proportion of cells in $M$ equal to 1) corresponding to respective plots (a) and (b).

% Wrap up
It is thus reassuring that all transmissions in $T$ can be identified at the receiver (TP = 100\%) even on the most challenging channel loads.
However, this comes at the expense of an increasingly large number of incorrect frames (FP) that must be filtered out at the gateway.

\subsection{Extraction Effectiveness} 
\label{sec:extraction}

Another measure of effectiveness is the number of actual transmissions that can be recovered at the gateway, also called the extraction probability. 
We consider two distinct cases in this section:
\begin{itemize}
    \item Headerfull case: at least one header replica has to be received, and a sufficient proportion of the fragments must be received without collision. This is how \lr is specified. Data is obtained from the model in Section~\ref{section:motivation} (complete frames in Figure~\ref{fig:extraction}).
    \item Headerless case: no header replica reception is required (thanks to our algorithms achieving 100\% TP), and a sufficient proportion of the fragments must be received without collision. Data is obtained from the simulation campaign introduced in Section~\ref{sec:detection} for the ILP model and the heuristic.
\end{itemize}

Figure~\ref{fig:fig5} shows the extraction probability as a function of the number of frame transmissions for the fast configuration (see Figure~\ref{fig:fig5}(a)) and the robust configuration (see Figure~\ref{fig:fig5}(b)).
For both configurations and for all values of the number of fragments per frame, the gain brought by the ILP or heuristic algorithms (termed headerless) over the legacy \lr is significant. For instance, in the fast configuration and for 2000 frame transmissions, the extraction ratio is 1\% for legacy \lr and 30\% for our heuristic. This shows that decoding headerless \lr frames can bring huge gains.

% Efficiency
\subsection{Computation Efficiency}

Figure~\ref{fig:ctime} indicates the computation time required to execute the ILP model and heuristic algorithm in practice.

% Metrics and Hardware
The primary metric we analyze is the compute (wall-clock) time that spans the execution of each solution.
To this end, we executed a series of benchmarking runs on the same hardware.
Specifically, we leveraged a WSL2 system running Python 3.8.10 and Gurobi 10.0.0 on an AMD Ryzen 9 5900HS clocking at 3.30 GHz with 16 GB of RAM.

% Plots
To ease the comparison, Figure~\ref{fig:ctime} presents the compute time for the same scenarios as in the previous sub-section (parameters in Table~\ref{table:params}).
We present the compute time for both the ILP model (dashed lines) and the heuristic algorithm (solid lines) as a function of the number of frame transmissions (Figure~\ref{fig:ctime}(a)) and total fragment count (Figure~\ref{fig:ctime}(b)).

% Observation 1: general
The first observation is that the ILP model can take up to 40 seconds in compute time to solve the most challenging cases with 1000 time slots and 3300 concurrent transmissions.
However, as expected, the heuristic algorithm delivers the same effectiveness with 17 seconds in the worst case. 
Therefore, we observe that the algorithmic solution is 2.3 times more efficient than state-of-the-art solvers.

% Observation 2: fragments
The second observation is that the compute time scales primarily with the fragment count rather than the transmitted frames.
This is reasonable, considering that the combinatorial problem complexity depends on the number of occupied slots in $M$.
However, we observe a secondary (weaker) correlation with the number of fragments per frame.
A larger number of fragments per frame seems to create more challenging problems for the same number of fragments (but also higher effectiveness, as discussed in the previous section).

\section{Discussion}
\label{section:discussion}
In this section, we first discuss the limits of the proposed solutions considering the assumptions we made. Then, we propose a way to mitigate false positives.

\subsection{Model Assumptions}
\label{subsection:discussion-assumptions}

\subsubsection{Known Fragment Count}
In \lr, the fragment count of a frame is stored in each replica of the header. In this study, we assumed that the number of fragments $\mathcal{P}$ is known and constant for each frame. Thus, this number does not have to be guessed when all replicas are lost. However, \lr is robust when handling frames of under-estimated (or even over-estimated) length, as the payload can be decoded even if 33\% (CR=2/3) or 67\% (CR=1/3) of the fragments are lost. We distinguish two cases:
\begin{enumerate}
    \item The identified sequence is too short: as long as a sufficient proportion of the first fragments are received and identified, the frame can still be decoded.
    \item The identified sequence is too long: the trailing fragments will be ignored, and the frame will be decoded without any problems.
\end{enumerate}

This assumption simplifies our model and keeps it linear.
Removing this assumption would require the model and the heuristic to deduce the fragment count. For the heuristic, this requires stopping the innermost {\tt for} loop (see line 5 of Algorithm~\ref{alg:heuristic}) whenever the channel is detected free. For the ILP model, a linearization strategy is needed for the term $\mathcal{P} \cdot y_{t,s}$ which would involve the product of a binary variable and a bounded variable in $\mathbb{N}$.
% However, LR-FHSS is robust when handling frames of under-estimated (or even over-estimated) length, as the payload can be decoded even if 33\% (CR=2/3) or 67\% (CR=1/3) of the fragments are lost. Thus, as long as a sufficient proportion of the first fragments are received, the trailing fragments can be ignored.

%However, we believe that we could weaken this assumption by solving a non-linear model and modifying the heuristic algorithm to deal with transmissions of unknown lengths. This improvement is attractive to future research emerging from this work.
% thanks to the high reliability of \lr coding rate (which can handle 33\% or 67\% fragment loss), -> why this would weaken the assumption, Alex? you can bring it back in with more explanation

\subsubsection{No Capture Effect and Good Fragment Detection}

We did not consider the capture effect: when two fragments collide, we consider that none can be retrieved. This is particularly relevant for the extraction effectiveness results summarized in Figure~\ref{fig:fig5}. In practice, however, likely, a transmission with a significantly larger signal strength than a weak colliding transmission will be retrievable, thus improving the number of recoverable frames compared to our results. Determining the actual capture effect threshold requires further investigation of the \lr modulation.

We also assumed that fragments can always be detected (but not recovered in case of collision) at the gateway, even if they collide. For this assumption, we rely on the good robustness of the GMSK modulation, which is the underlying modulation for the fragments~\cite{lebreton22crash}. For the heuristic, bad fragment detection (including due to noise) can be taken into account by adding $(s,t,\mathcal{P})$ in the set $\hat{T}$ if a sufficient proportion of the corresponding OBWs are detected busy, rather than all corresponding OBWs.

\subsubsection{Time Slotting}

Our strongest assumption is that transmissions are slotted at the fragment level, that is, every 102.4~ms. Transmissions will not be slotted in practice, and our algorithms must deal with arbitrary transmission times.

When a transmission is not aligned with the slots, each fragment will overlap with two slots. The receiving energy of the fragment is distributed into the two slots based on the overlap time. Suppose the overlap between an unaligned fragment and a slot is small. In that case, the energy of this fragment on the slot becomes negligible, and our assumption applies as the fragments are nearly aligned. If the overlap between an unaligned fragment and a slot is large (up to half of the duration of the slot), the model could detect the residual energy during two consecutive slots and incorrectly assume that transmission occurred during these two slots. Dealing with arbitrary transmission times remains challenging and requires the research on \lr to progress on characterizing its capture effect capabilities and its tolerance to desynchronization.

\subsection{False Positive Filtering}

The two algorithms produce many FP when the channel is heavily loaded. 
The gateway will consider each FP as a frame with an arbitrary payload. Two practical strategies can mitigate the impact of these fake frames. 

On the one hand, it is possible to use the cyclic redundancy check (CRC) present in the LoRaWAN frame to check the \textit{validity} of the received payload. Each LoRaWAN frame contains a CRC calculated on the whole payload. If the CRC value in the received frame differs from the CRC computed on the decoded payload, the gateway will drop the frame. Thus, many fake FPs will be removed due to incorrect CRC checks at the gateway. With a 16-bit CRC, on average, only one FP every $2^{16}$ would still be accepted. For one million FP, the average number of \textit{fake} frames would be about 15.

On the other hand, FP occurs when a sequence fully overlaps with existing transmissions. If this occurs, it means that most of the sequence slots will collide, and thus the gateway will not be able to retrieve actual data from the colliding fragments, and the sequence will not translate into a valid frame. Thus, it is likely that FP will produce corrupted (fake) frames.

\section{Conclusion}
\label{section:conclusion}

In this paper, we aim to improve \lr throughput by enabling the recovery of frames whose header was lost. First, we showed that header loss is a frequent cause of frame loss in \lr. Then, we proposed an exact ILP and a low-cost heuristic to detect and extract headerless frames profiting from the \textit{a priori} known sequence set. We showed through extensive simulations that our heuristic consistently achieves optimal results in many configurations while incurring tractable computation costs. Thus, we believe that our heuristic can successfully be implemented on \lr gateways located onboard of LEO satellite and help deal with the massive amount of incoming concurrent transmissions. Our future work involves weakening some of the assumptions of this initial work, including a more realistic channel model and transmission schedule.

This work encourages more experimental work on \lr to help improve the current models, \eg, giving thresholds for capture effect.

%Future work involves
%a) fragment loss model (signal loss due to weak power),
%b) collision-aware model (number of overlapping fragments is available)
%c) non-synchronized simulation analysis,
%d) interaction with frames with headers and other transmissions (e.g., legacy LoRa),
%e) model for unknown frame size 

\balance
\bibliographystyle{abbrv}
\bibliography{references}  % sigproc.bib is the name of the Bibliography in this case

\end{document}